\documentclass[runningheads]{llncs}
\usepackage[T1]{fontenc}
\usepackage{graphicx}
\usepackage{tabularray}
\usepackage{hyperref}
\begin{document}
\sloppy
\title{Unlearning Spurious Correlations in Chest X-ray Classification}

%in Chest X-ray Classification\thanks{Supported by organization x.}


\author{Misgina Tsighe Hagos\inst{1,2}\orcidID{0000-0002-9318-9417} \and
Kathleen M. Curran\inst{1,3}\orcidID{0000-0003-0095-9337} \and
Brian Mac Namee\inst{1,2}\orcidID{0000-0003-2518-0274}}
\authorrunning{M. T Hagos et al.}
\institute{Science Foundation Ireland Centre for Research Training in Machine Learning\\ 
\email{misgina.hagos@ucdconnect.ie}\\
%\url{http://www.springer.com/gp/computer-science/lncs} 
\and
School of Computer Science, University College Dublin, Ireland\\
\email{brian.macnamee@ucd.ie}\\
\and
School of Medicine, University College Dublin, Ireland\\
\email{kathleen.curran@ucd.ie}}

\maketitle              % typeset the header of the contribution
\begin{abstract}
Medical image classification models are frequently trained using training datasets derived from multiple data sources. While leveraging multiple data sources is crucial for achieving model generalization, it is important to acknowledge that the diverse nature of these sources inherently introduces unintended confounders and other challenges that can impact both model accuracy and transparency. A notable confounding factor in medical image classification, particularly in musculoskeletal image classification, is skeletal maturation-induced bone growth observed during adolescence. We train a deep learning model using a Covid-19 chest X-ray dataset and we showcase how this dataset can lead to spurious correlations due to unintended confounding regions. eXplanation Based Learning (XBL) is a deep learning approach that goes beyond interpretability by utilizing model explanations to interactively unlearn spurious correlations. This is achieved by integrating interactive user feedback, specifically feature annotations. In our study, we employed two non-demanding manual feedback mechanisms to implement an XBL-based approach for effectively eliminating these spurious correlations. Our results underscore the promising potential of XBL in constructing robust models even in the presence of confounding factors.

% , which was recognized as the winner of the Covid-19 dataset award from the Kaggle community. Our objective is to
% The abstract should briefly summarize the contents of the paper in

\keywords{Interactive Machine Learning \and eXplanation Based Learning \and Medical Image Classification \and Chest X-ray}
\end{abstract}

\section{Introduction}
\label{section:introduction}

While Computer-Assisted Diagnosis (CAD) holds promise in terms of cost and time savings, the performance of models trained on datasets with undetected biases is compromised when applied to new and external datasets. This limitation hinders the widespread adoption of CAD in clinical practice \cite{zech2018variable,santa2021public}. Therefore, it is crucial to identify biases within training datasets and mitigate their impact on trained models to ensure model effectiveness. 

For example, when building models for the differential diagnosis of pathology on chest X-rays (CXR) it is important to consider skeletal growth or ageing as a confounding factor. This factor can introduce bias into the dataset and potentially mislead trained models to prioritize age classification instead of accurately distinguishing between specific pathologies. The effect of skeletal growth on the appearance of bones necessitates careful consideration to ensure that a model focuses on the intended classification task rather than being influenced by age-related features. 

\begin{figure}[t]
\vskip 0.2in
\begin{center}
\centerline{\includegraphics[width=0.6\linewidth]{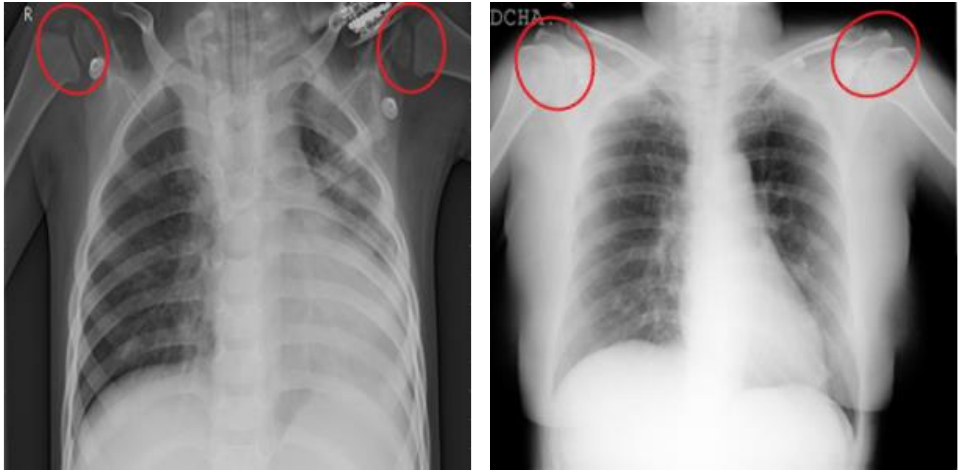}}
\vskip -0.08in
\caption{In the left image, representing a child diagnosed with Viral pneumonia, the presence of Epiphyses on the humerus heads is evident, highlighted with red ellipses. Conversely, the right image portrays an adult patient with Covid-19, where the Epiphyses are replaced by Metaphyses, also highlighted with red ellipses.} % The comparison of two chest X-ray images showcases the anatomical differences between a child (left image) and an adult participant (right image). 
\label{figure:pneumonia_vs_covid}
\end{center}
\vskip -0.3in
\end{figure}

An illustrative example of this scenario can be found in a recent study by Pfeuffer et al. \cite{pfeuffer2023explanatory}. In their research, they utilized the Covid-19 CXR dataset \cite{chowdhury2020can}, which includes a category comprising CXR images of children. This dataset serves as a pertinent example to demonstrate the potential influence of age-related confounders, given the presence of images from pediatric patients. It comprises CXR images  categorized into four groups: Normal, Covid, Lung opacity, and Viral pneumonia. However, a notable bias is introduced into the dataset due to the specific inclusion of the Viral pneumonia cases collected exclusively from children aged one to five years old \cite{kermany2018identifying}. This is illustrated in Figure \ref{figure:pneumonia_vs_covid} where confounding regions introduced due to anatomical differences between a child and an adult in CXR images are highlighted. Notably, the presence of Epiphyses in images from the Viral pneumonia category (which are all from children) is a confounding factor, as it is not inherently associated with the disease but can potentially mislead a model into erroneously associating it with the category. Addressing these anatomical differences is crucial to mitigate potential bias and ensure accurate analysis and classification in pediatric and adult populations. 

Biases like this one pose a challenge to constructing transparent and robust models capable of avoiding spurious correlations. Spurious correlations refer to image regions that are mistakenly believed by the model to be associated with a specific category, despite lacking a genuine association. % Overcoming this bias and ensuring accurate classification necessitate careful consideration of these confounding factors and the development of robust learning approaches.
% The variations in size, shape, and development of anatomical structures, such as the heart, lungs, and bones, between children and adults can introduce confounding factors that affect the interpretation and classification of the X-ray images.

% \footnote{https://www.kaggle.com/datasets/tawsifurrahman/covid19-radiography-database}

% one to five years old\footnote{https://www.kaggle.com/datasets/paultimothymooney/chest-xray-pneumonia}

 % from a paediatric clinic at Guangzhou Women and Children’s Medical Centre \cite{kermany2018identifying}. % 

% Even though the Covid-19 Radiography dataset is an important contribution for CAD research, the chest x-ray images are biased by age. 

While the exact extent of affected images remains unknown, it is important to note that the dataset also encompasses other confounding regions, such as texts and timestamps. However, it is worth mentioning that these confounding regions are uniformly present across all categories, indicating that their impact is consistent throughout. For the purpose of this study, we specifically concentrate on understanding and mitigating the influence of musculoskeletal age in the dataset. % addressing the bias caused by age-related factors. %Although other confounding regions exist, our focus is primarily directed towards understanding and mitigating the influence of musculoskeletal age in the dataset. 

% For this reason, we believe these confounding image regions are not as challenging and don't cause bias as much as the age difference of participants between the categories and we focus only on the age-caused bias.

eXplanation Based Learning (XBL) represents a branch of Interactive Machine Learning (IML) that incorporates user feedback in the form of feature annotation during the training process to mitigate the influence of confounding regions \cite{schramowski2020making}. By integrating user feedback into the training loop, XBL enables the model to progressively improve its performance and enhance its ability to differentiate between relevant and confounding features \cite{hagos2022identifying}. In addition to unlearning spurious correlations, XBL has the potential to enhance users' trust in a model  \cite{dietvorst2018overcoming}. By actively engaging users and incorporating their expertise, XBL promotes a collaborative learning environment, leading to increased trust in the model's outputs. This enhanced trust is crucial for the adoption and acceptance of models in real-world applications, particularly in domains where decisions have significant consequences, such as medical diagnosis. 

XBL approaches typically add  regularization to the loss function used when training a model, enabling it to disregard the impact of confounding regions. A typical XBL loss can be expressed as:

% User feedback, collected through model explanations, plays a vital role in guiding the learning process and refining the model. 

% By involving users in the training process and providing them with explanations of model decisions, XBL fosters transparency and interpretability. This interactivity allows users to gain insights into how the model arrives at its predictions, enabling them to provide feedback and contribute to model refinement.

\begin{equation}
\label{formula:generic_xbl}
    L = L_{CE} + L_{expl} + \lambda{}\sum_{i=0}\theta_i^2  \hspace{0.1cm},
\end{equation}
% \begin{equation}
% \label{equation:schramowski_gradcam}
%     L_{expl} = \sum_{i=0}^N M_{i}GradCAM(x_{i})
% \end{equation}

\noindent where $L_{CE}$ is categorical cross entropy loss that measures the discrepancy between the model's predictions and ground-truth labels; $\lambda$ is a regularization term; $\theta$ refers to network parameters; and $L_{expl}$ is an explanation loss. Explanation loss can be formulated as:

\begin{equation}
\label{formula:exp_loss}
    L_{expl} = \sum_{i=0}^N M_{i} \odot Exp(x_{i})  \hspace{0.1cm},
\end{equation}

\noindent where $N$ is the number of training instances, $x \in X$; $M_i$ is a manual annotation of confounding regions in the input instance $x_i$; and $Exp(x_{i})$ is a saliency-based model explanation for instance $x_i$, for example generated using Gradient weighted Class Activation Mapping (GradCAM) \cite{schramowski2020making}. GradCAM is a feature attribution based model explanation that computes the attention of the learner model on different regions of an input image, indicating the regions that significantly contribute to the model's predictions \cite{selvaraju2017grad}. This attention serves as a measure of the model's reliance on these regions when making predictions. The loss function, $L_{expl}$, is designed to increase as the learner's attention to the confounding regions increases. Overall, by leveraging GradCAM-based attention and the associated $L_{expl}$ loss, XBL provides a mechanism for reducing a model's attention to confounding regions, enhancing the interpretability and transparency of a model's predictions.  %This attention is computed using GradCAM.

\begin{figure}[t]
\vskip 0.2in
\begin{center}
\centerline{\includegraphics[width=0.75\linewidth]{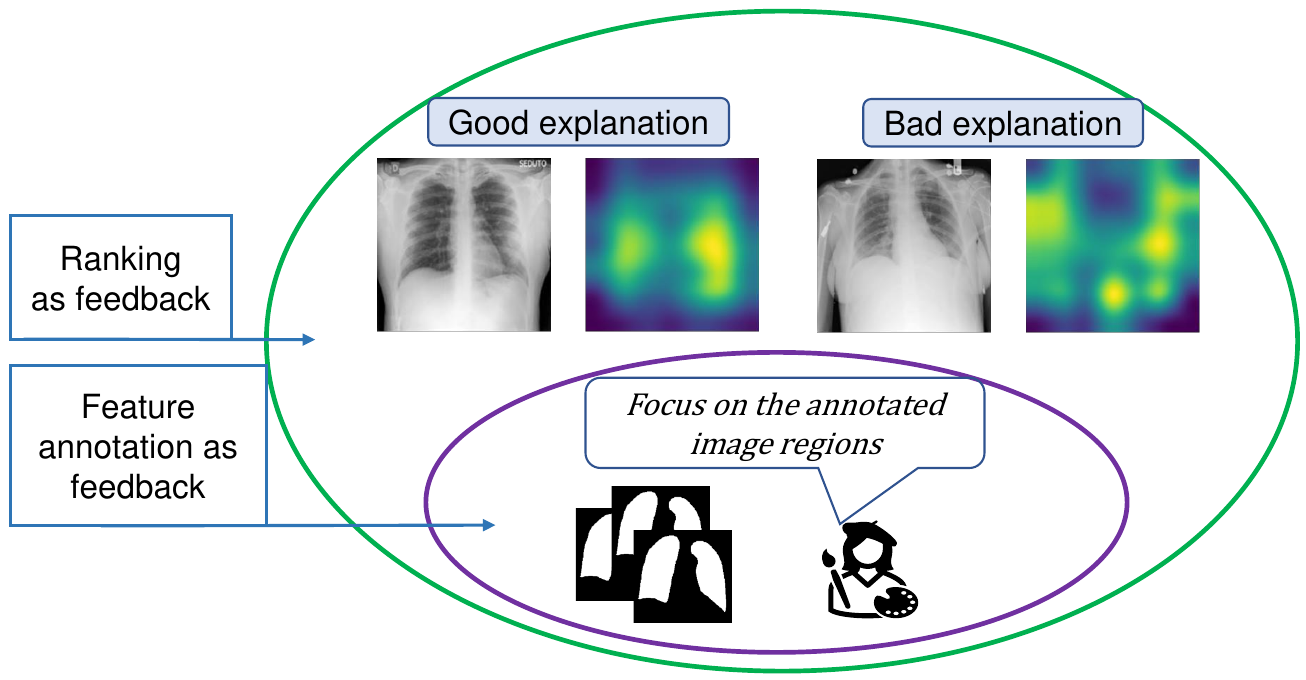}}
\vskip -0.08in
\caption{The inner ellipse shows the typical mode of feedback collection where users annotate image features. The outer ellipse shows how our proposed approach requires only identification of one good and one bad explanation.}
\label{figure:proposed_method}
\end{center}
\vskip -0.2in
\end{figure}

As is seen in the inner ellipse of Figure \ref{figure:proposed_method}, in XBL, the most common mode of user interaction is image feature annotation. This requires user engagement that is considerably  more demanding than the simple instance labeling that most IML techniques require \cite{zlateski2018importance} and increases the time and cost of feedback collection. As can be seen in the outer ellipse of Figure \ref{figure:proposed_method}, we are interested in lifting this pressure from users (feedback providers) and simplifying the interaction to ask for identification of two explanations as exemplary explanations and ranking them as good and bad explanations.  This makes collecting feedback  cheaper and faster. This kind of user interaction where users are asked for a ranking instead of category labels has also been found to increase inter-rater reliability and data collection efficiency \cite{o2017rating}. We incorporate this feedback into model training through a contrastive triplet loss \cite{chechik2010large}.

% In this paper, we build two simple variants of XBL that are based on only two feedbacks, and we show that XBL can be implemented to remove spurious correlations and to build robust models.
The main contributions of this paper are:
\begin{enumerate}
    
    \item We propose the first type of eXplanation Based Learning (XBL) that can learn from only two exemplary explanations of two training images; % that can train robust models using only two exemplary explanations of two training images;
    \item We present an approach to adopt triplet loss for XBL to incorporate the two exemplary explanations into an explanation loss;
    \item Our experiments demonstrate that the proposed method achieves improved explanations and comparable classification performance when compared against a baseline model. % In addition to showing that XBL can be implemented with limited user feedback to remove spurious correlations, o
\end{enumerate}

\section{Related Work}
\label{section:related_work}

\subsection{Chest x-ray classification}

A number of Covid-19 related datasets have been collated and deep learning based diagnosis solutions have been proposed due to the health emergency caused by Covid-19 and due to an urgent need for computer-aided diagnosis (CAD) of the disease \cite{islam2021review}. In addition to training deep learning models from scratch, transfer learning, where parameters of a pre-trained model are further trained to identify Covid-19, have been utilized \cite{yousefzadeh2021ai}. Even though the array of datasets and deep learning models show promise in implementing CAD, care needs to be taken when the datasets are sourced from multiple imaging centers and/or the models are only validated on internal datasets. The Covid-19 CXR dataset, for example, has six sources at the time of writing this paper. This can result in unintended confounding regions in images in the dataset and subsequently spurious correlations in trained models \cite{santa2021public}.
% aljondi2020diagnostic,almeida2020lung
% Ultrasound, Computer Tomography (CT), and chest x-ray modalities have been used to collect datasets for the training of Covid-19 diagnosis models, with x-rays being the easiest to apply and widely available \cite{aljondi2020diagnostic}.

% n utilised \cite{wu2020deep,yousefzadeh2021ai}.

% and one of the imaging centres only collected x-ray images from children aged one to five

\subsection{eXplanation Based Learning}

XBL can generally be categorized based on how feedback is used: (1) augmenting loss functions; and (2) augmenting training datasets.

\paragraph{Augmenting Loss Functions.}

As shown in Equation \ref{formula:generic_xbl}, approaches in this category  add an explanation loss, $L_{expl}$, during model training to encourage focus on image regions that are considered relevant by user(s), or to ignore confounding regions \cite{hagos2022impact}. Ross et al. \cite{ross2017right} use an $L_{expl}$ that penalizes a model with high input gradient model explanations on the wrong image regions based on user annotation, 

% $L_{expl} = \sum_{n}^N \left[ M_n \cdot \frac{\partial} {\partial x_n}\sum_{k=1}^K \log{\hat{y}_{nk}} \right] ^2$

\begin{equation}
\label{equation:rrr}
    L_{expl} = \sum_{n}^N \left[ M_n \odot \frac{\partial} {\partial x_n}\sum_{k=1}^K \log{\hat{y}_{nk}} \right] ^2   ,
\end{equation}  

\noindent for a function $f(X|\theta)=\hat{y} \in {R}^{N\times K} $ trained on $N$ images, $x_{n}$, with $K$ categories, where $M_{n} \in \ \{0,\ 1\}$ is user annotation of confounding image regions. Similarly, Shao et al. \cite{shao2021right} use influence functions in place of input gradients to correct a model's behavior  % One example is Right for the Right Reasons (RRR) that penalises a model with high input gradient model explanations on the wrong image regions based on user annotation \cite{ross2017right}. 

% \noindent for a function $f(X|\theta)=\hat{y} \in {R}^{N\times K} $ trained on $N$ images, $x_{n}$, with $K$ categories, where $M_{n} \in \ \{0,\ 1\}$ is user annotation of confounding image regions. Similarly, Shao et al. (2021) use Influence Functions (IF) in place of input gradients to correct a model's behaviour \cite{shao2021right}. % Contextual Decomposition Explanation Penalisation (CDEP) \cite{rieger2020interpretations} penalises features and feature interactions.

% A  GradCAM model explanation was used instead of input gradients in RRR-G by Schramowski \textit{et al.} \cite{schramowski2020making} using the following loss function:

%\begin{eqnarray}
%     L_{expl} = \sum_{n}^N M_{n}GradCAM(x_{n})
% \end{eqnarray}
% \begin{math}\end{math}. 

% User feedback in XBL experiments can be either: (1) telling the model to ignore non-salient image regions; or (2) instructing the model to focus on important image regions in a training dataset \cite{hagos2022impact}. While the XBL methods presented above refine a model by using the first feedback type, Human Importance-aware Network Tuning (HINT) does the opposite by teaching a model to focus on important image parts using GradCAM model explanations \cite{selvaraju2019taking}.

\paragraph{Augmenting Training Dataset.} In this category, a confounder-free dataset is added to an existing confounded training dataset to train models to avoid learning spurious correlations. In order to unlearn spurious correlations from a classifier that was trained on the Covid-19 dataset, Pfeuffer et al. \cite{pfeuffer2023explanatory} collected feature annotation on 3,000 chest x-ray images and augmented their training dataset. This approach, however, doesn't  target unlearning or removing  spurious correlations, but rather adds a new variety of data. This means models are being trained on a combination of the existing confounded training dataset and the their new dataset. %In addition, this kind of task hinders practical deployment and domain transferability of XBL because it requires user engagement for the demanding task of feature annotation of training datasets for every domain area of application \cite{zlateski2018importance}. 

%While XBL approaches show promise in unlearning spurious correlations \cite{hagos2022identifying,pfeuffer2023explanatory}, they all need a lot of effort from users. 

One thing all approaches to XBL described above have in common is the assumption that users will provide feature annotation for all  training instances to refine or train a model. We believe that this level of user engagement hinders practical deployment of XBL because of the demanding nature and expense of feature annotation that is required \cite{zlateski2018importance}. It is, therefore, important to build an XBL method that can refine a trained model using a limited amount of user interaction and we propose eXemplary eXplanation Based Learning to achieve this.

%Instance relabelling \cite{teso2021interactive}, counterexamples generation \cite{teso2019explanatory}, and using user feedback as new training instances \cite{popordanoska2020machine} are some of the methods that augment a dataset to incorporate user feedback into XBL.

% Instance relabelling has been deployed to clean label noise in a training dataset that is identified using example based explanations \cite{teso2021interactive}. Counter-Examples (CE), which are variants of training instances with added modifications using user feedback can be generated to augment dataset for model re-training \cite{teso2019explanatory}. Simpler surrogate models have also been used as global explanations to elicit feedback in the form of new training instances \cite{popordanoska2020machine}.

% Although they only considered three categories, in the literature, Pfeuffer et al. \yrcite{pfeuffer2023explanatory} is the closest to our work.

\section{eXemplary eXplanation Based Learning}
\label{section:eXBL}

User annotation of image features, or $M$, is an important prerequisite for typical XBL approaches (illustrated in Equation \ref{formula:generic_xbl}). We use eXemplary eXplanation Based Learning (eXBL) to reduce the time and resource complexity caused by the need for $M$. eXBL simplifies the expensive feature annotation requirement by replacing it with identification of just two exemplary explanations: a \emph{Good explanation} ($C_{good_{i}}$) and a \emph{Bad explanation} ($C_{bad_{j}}$) of two different instances, $x_{i}$ and $x_{j}$. We pick the two exemplary explanations manually based on how much attention a model's explanation output gives to relevant image regions. A good explanation would be one that gives more focus to the lung and chest area rather than the irrelevant regions such as the Epiphyses, humerus head, and image backgrounds, while a bad explanation does the opposite. %However, even if this replaces feature annotation with two labels, categorizing explanations would still be expensive if it's to be performed for all training instances whose size could be in the thousands. For this reason, we only use one $C_{good_{i}}$ and one $C_{bad_{j}}$. %a single instance GradCAM explanation for each category.

We choose to use GradCAM model explanations because they have been found to be more sensitive to training label reshuffling and model parameter randomization than other saliency based explanations \cite{adebayo2018sanity}; and they provide accurate explanations in medical image classifications \cite{marmolejo2022numerical}. 

% \begin{figure}[t]
% \vskip 0.2in
% \begin{center}
% \centerline{\includegraphics[width=0.6\linewidth]{good_bad_explanations.pdf}}
% \caption{Good (top row) and bad (bottom row) Grad-CAM explanations; [MIDDLE] GradCAM; [RIGHT] GradCAM overlaid over input images in the left.}
% \label{figure:good_bad_explanations}
% \end{center}
% \vskip -0.3in
% \end{figure}

% We then assign products of input instances and their corresponding selected GradCAM explanations to $C_{good}$ and $C_{bad}$,

We then compute product of the input instances and the Grad-CAM explanation in order to propagate input image information towards computing the loss and to avoid a bias that may be caused by only using a model's GradCAM explanation,

\begin{equation}
    C_{good} := x_{i}\odot C_{good_{i}} %GradCAM(i) 
\end{equation}
\begin{equation}
    C_{bad} := x_{j}\odot C_{bad_{j}} %GradCAM(j)
\end{equation}

% The product of the input instance and the Grad-CAM explanation is used instead of just the Grad-CAM explanation to propagate input image information and because taking the GradCAM outputs only could lead to biased exemplary explanations as we would be only taking the model's attention into consideration. 

% For this reason, We take the product of instances $i$ and $j$ against the GradCAM explanations to propagate and make use of the information of the input instances in computing $C_{good}$ and $C_{bad}$ in addition to the model's attention.

We then take inspiration from triplet loss \cite{chechik2010large} to incorporate $C_{good}$ and $C_{bad}$ into our explanation loss, $L_{expl}$. The main purpose of $L_{expl}$ is to penalize a trainer according to similarity of model explanations of instance $x$ to $C_{good}$ and its difference from $C_{bad}$. We use Euclidean distance as a loss to compute the measure of dissimilarity, $d$ (loss decreases as similarity to $C_{good}$ is high and to $C_{bad}$ is low). 

% For the product of the training instances $x\in X$, and their corresponding GradCAM outputs, $x \cdot GradCAM(x)$, we compute the similarity function $S_{xg}$ and $S_{xb}$, which represent distances from $C_{good}$ and $C_{bad}$ as follows,

\begin{equation}
\label{equation:dxg}
    d_{xg} := d(x \odot GradCAM(x), C_{good})
\end{equation}
\begin{equation}
\label{equation:dxb}
    d_{xb} := d(x \odot GradCAM(x), C_{bad})
\end{equation}

We train the model $f$ to achieve $d_{xg} \ll d_{xb}$ for all $x$. We do this by adding a $margin = 1.0$ and translating it to:  $d_{xg} < d_{xb} + margin$. We then compute the explanation loss as:
% so $GradCAM(x)$ can be aligned to resemble the good explanation

\begin{equation}
    \label{equation:proposed_lexp}
    L_{expl} = \sum_{i}^N \max(d_{x_{i}g} - d_{x_{i}b} + margin, 0) 
\end{equation}

In addition to correctly classifying $X$, which is achieved through $L_{CE}$, this $L_{expl}$ (Equation \ref{equation:proposed_lexp})  trains $f$ to output GradCAM values that resemble the good explanations and that differ from the bad explanations, thereby refining the model to focus on the relevant regions and to ignore confounding regions. $L_{expl}$ is zero, for a given sample $x$, unless $x \odot GradCAM(x)$ is much more similar to $C_{bad}$ than it is to $C_{good}$---meaning $d_{xg} > d_{xb} + margin$.

\section{Experiments}
\label{section:experiments}

\subsection{Data Collection and Preparation}

To demonstrate eXBL we use the Covid-19 CXR dataset \cite{chowdhury2020can,rahman2021exploring} described in Section \ref{section:introduction}. For model training we subsample 800 x-ray images per category to mitigate class imbalance, totaling 3,200 images. For validation and testing, we use 1,200 and 800  images respectively. We resize all images to 224 $\times$ 224 pixels. The dataset is also accompanied with feature annotation masks that show the lungs in each of the x-ray images collected from radiologists \cite{rahman2021exploring}.

% Even though the exact number of effected images is unknown, the dataset contains confounding regions, such as marks, texts, and timestamps in many of the images.

\subsection{Model Training}

We followed a transfer learning approach using a pre-trained MobileNetV2 model \cite{sandler2018mobilenetv2}. We chose to use MobileNetV2 because it achieved better performance at the CXR images classification task at a reduced computational cost after comparison among pre-trained models. In order for the training process to affect the GradCAM explanation outputs, we only freeze and reuse the first 50 layers of MobileNetV2 and retrain the rest of the convolutional layers with a custom classifier layer that we added (256 nodes with a ReLu activation with a 50\% dropout followed by a Softmax layer with 4 nodes).

We first trained the MobileNetV2 to categorize the training set into the four classes using categorical cross entropy loss. It was trained for 60 epochs\footnote{The model was trained with an early stop monitoring the validation loss at a patience of five epochs and a decaying learning rate = 1e-04 using an Adam optimizer.}. We refer to this model as the Unrefined model. We then use the Unrefined model to select the good and bad explanations displayed in Figure \ref{figure:proposed_method}. Next, we employ our eXBL algorithm using the good and bad explanations to teach the Unrefined model to focus on relevant image regions by tuning its explanations to look like the good explanations and to differ from the bad explanations as much as possible. We use Euclidean distance to compute dissimilarity in adopting a version of the triplet loss for XBL. We refer to this model as the eXBL$_{EUC}$ model and it was trained for 100 epochs using the same early stopping, learning rate, and optimizer as the Unrefined model.

For model evaluation, in addition to classification performance, we compute an objective explanation evaluation using Activation Precision \cite{barnett2021case} that measures how many of the pixels predicted as relevant by a model are actually relevant using existing feature annotation of the lungs in the employed dataset, %, $AP = \frac{1}{N} \sum_{n}^N\frac{ GradCAM_{\theta}(x_{n}) * A_{obj_{n}}}{GradCAM_{\theta}(x_{n})}$.

\begin{equation}
\label{formula:activation_precision}
    AP = \frac{1}{N} \sum_{n}^N\frac{ \sum (T_{\tau}(GradCAM_{\theta}(x_{n})) \odot A_{x_{n}})}{\sum(T_{\tau}(GradCAM_{\theta}(x_{n})))} \hspace{0.1cm},
\end{equation} 

\noindent where $x_{n}$ is a test instance, $A_{x_{n}}$ is feature annotation of lungs in the dataset, $GradCAM_{\theta}(x_{n})$ holds the GradCAM explanation of $x_{n}$ generated from a trained model, and $T_{\tau}$ is a threshold function that finds the (100-$\tau$) percentile value and sets elements of the explanation, $GradCAM_{\theta}(x_{n})$, below this value to zero and the remaining elements to one. In our experiments, we use $\tau=5\%$.

\begin{figure}[!h]
\vskip 0.2in
\begin{center}
\centerline{\includegraphics[width=1\linewidth]{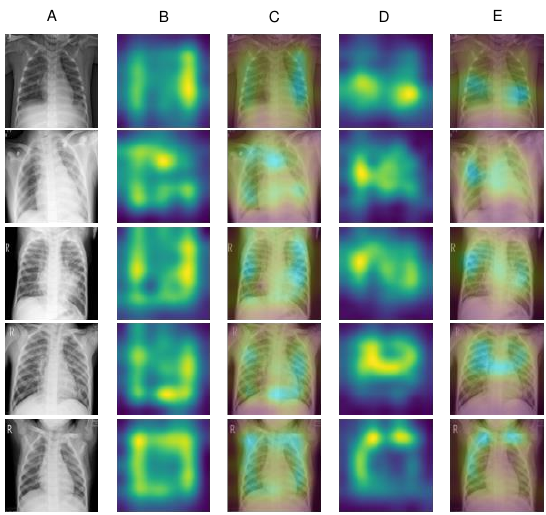}}
% \vskip -0.06in
\caption{Sample outputs of Viral Pneumonia category. (A) Input images; (B) GradCAM outputs for Unrefined model and (C) their overlay over input images; (D) GradCAM outputs for eXBL$_{EUC}$ and (E) their overlay over input images.}
\label{figure:sample_gradcam_explanations}
\end{center}
\vskip -0.2in
\end{figure}

\section{Results}
\label{section:results}

Table \ref{table:summary_performance} shows classification and explanation performance of the Unrefined and eXBL$_{EUC}$ models. Sample test images, GradCAM outputs, and overlaid GradCAM visualizations of x-ray images with Viral pneumonia category are displayed in Figure \ref{figure:sample_gradcam_explanations}. From the sample GradCAM outputs and Table \ref{table:summary_performance}, we observe that the eXBL$_{EUC}$ model was able to produce more accurate explanations that avoid focusing on irrelevant image regions such as the Epiphyses and background regions. This is demonstrated by how GradCAM explanations of the eXBL$_{EUC}$ model tend to focus on the central image regions of the input images focusing around the chest that is relevant for the classification task, while the GradCAM explanations generated using the Unrefined model give too much attention to areas around the shoulder joint (humerus head) and appear angular shaped giving attention to areas that are not related with the disease categories. % However, the superior explanations of the XBL models come with a classification performance loss as is summarised in Table \ref{table:summary_performance}. 
% eXBL outperforms the Unrefined model in explanation performance, meaning in correctly locating and using the relevant image regions for classification. 

% \usepackage{tabularray}
\begin{table}[t]
\vskip -0.1in
\centering
\caption{Classification and explanation performance.}
\label{table:summary_performance}
\begin{tblr}{
  row{2} = {c},
  cell{1}{1} = {r=2}{},
  cell{1}{2} = {c=2}{c},
  cell{1}{4} = {c=2}{},
  cell{3}{2} = {c},
  cell{3}{3} = {c},
  cell{3}{4} = {c},
  cell{3}{5} = {c},
  cell{4}{2} = {c},
  cell{4}{3} = {c},
  cell{4}{4} = {c},
  cell{4}{5} = {c},
  cell{5}{2} = {c},
  cell{5}{3} = {c},
  cell{5}{4} = {c},
  cell{5}{5} = {c},
  hline{1,3,6} = {-}{},
  hline{2} = {2-5}{},
}
Models    & Accuracy   &      & Activation Precision  &      \\
          & Validation & Test & Validation            & Test \\
Unrefined & 0.94       & 0.95 & 0.32                  & 0.32 \\
eXBL$_{EUC}$   & 0.89       & 0.90 & 0.34                  & 0.35
\end{tblr}
\vskip -0.3in
\end{table}

\section{Conclusion}
\label{section:discussion_and_conclusion}

In this work, we have presented an approach to debug a spurious correlation learned by a model and to remove it with just two exemplary explanations in eXBL$_{EUC}$. We present a way to adopt the triplet loss for unlearning spurious correlations. Our approach can tune a model's attention to focus on relevant image regions, thereby improving the saliency-based model explanations. We believe it could be easily adopted to other medical or non-medical datasets because it only needs two non-demanding exemplary explanations as user feedback.

Even though the eXBL$_{EUC}$ model achieved improved explanation performances when compared to the Unrefined model, we observed that there is a classification performance loss when retraining the Unrefined model with eXBL to produce good explanations. This could mean that the initial model was exploiting the confounding regions for better classification performance. It could also mean that our selection of good and bad explanations may not have been optimal and that the two exemplary explanations may be degrading model performance. 

% While subjective evaluations of model explanations could be used to collect users' view, we select to use an objective evaluation for a speedy assessment since our aim is to demonstrate capability of the proposed model training approach. In addition to the lung, the disease categories might be associated with other areas of the body such as the throat and torso \cite{pfeuffer2023explanatory}; however, we compute AP using lung annotations because lung is the only region of the chest x-rays annotated in the dataset.

% Since this work focuses on demonstrating the effectiveness of eXBL with limited and non-demanding user feedback, we depend on objective metrics for evaluation of the models' performance.

Since our main aim in this study was to demonstrate effectiveness of eXBL$_{EUC}$ based on just two ranked feedback, the generated explanations were evaluated using masks of lung because it is the only body part with pixel-level annotation in the employed dataset. However, in addition to the lung, the disease categories might be associated with other areas of the body such as the throat and torso. For this reason, and to ensure transparency in practical deployment of such systems in clinical practice, future work should involve expert end users for evaluation of the classification and model explanations. % Future work should compare effects of using more than two exemplary explanations on model performance. 

\section*{Acknowledgements}This publication has emanated from research conducted with the financial support of Science Foundation Ireland under Grant number 18/CRT/6183. For the purpose of Open Access, the author has applied a CC BY public copyright licence to any Author Accepted Manuscript version arising from this submission.

\bibliographystyle{splncs04}
\bibliography{splncs04}

\end{document}